\documentstyle{article}

\begin{document}

\title{A two-step model and the algorithm for\\
    recalling in recommender systems}

\author{Keisuke Hara
and Tomihisa Kamada \footnote{kshara@mynd.jp, tomy@mynd.jp}\\
Mynd, Inc.\\
Yoshida FG Bld.4F, 3-17 Kanda Jinbocho,\\
Chiyoda-ku, 101-0051 JAPAN}

\date{23 Oct. 2013}

\maketitle

\begin{abstract}

When a user finds an interesting recommendation
in a recommender system,
the user may want to recall related items recommended in the past
to reconsider or to enjoy them again.
If the system can pick up
such {\it recalled} items at each user's request,
it must deepen the user experience.

We propose a model and the algorithm for such personalized {\it recalling}
in conventional recommender systems,
which is an application of neural networks for associative memory.
In our model, the {\it recalled} items can reflect
each user's personality beyond naive similarities between items.

\end{abstract}

\section{Introduction}

We propose a new function to give {\it recalled} items
at each user's request in standard recommender systems.
For example, in a news recommender system,
when a user browses recommended news headlines
and finds an interesting one,
the user may remember articles recommended many days ago.
For the user, it might be crucial or simply enjoyable
to read them again.
If the system can pick up and show such {\it recalled} items
at each user's request,
it must give the users more convenience
and deepen the user experience.

In other saying,
our task is to choose items related to an image in each user's mind
stimulated by a given new item.
Therefore, the main difficulty
lies in the fact that this {\it recalling} process is personal
and that just similarities between items are not enough.

In a general sense, our {\it recalling} is a {\it personal search}.
There are many studies in the area of personalisation of search engines.
For example, it is an important issue
how to customise searched results based on each user's activity history
(\cite{Teevan}), or how to use user's memory to help the user's task
in computer systems (\cite{Brush}), etc.

Such general approaches, however, do not seem proper in our situation
since our {\it recalling} has it's own unique character and requirements.
Therefore, we should give a simple and effective
algorithm in our restricted situation, that is, in recommender systems.

To achieve this,
we propose a two-step model of each user's {\it recalling} process
in a recommender system.
First, for a given item,
we extend its features with the information of the old recommendations
to the user to construct the {\it recall vector}.
Next, we pick up {\it recalled} items that are near to the
{\it recall vector}.

Roughly saying,
the {\it recall vector}
is an association or a mental image in the user's mind
stimulated by the given item.
The essence of our algorithm is to bring
the idea of the classical neural networks for associative memory
developed by J.A.Anderson(\cite{Anderson}), T.Kohonen, K.Nakano
into the space of recommended items to each user.

In the following section,
we explain our {\it recalling} function
and give our model with the algorithm in an abstract way.
We discuss the problems in section 3
and summarise our results in the last section.

\section{A model and the algorithm for recalling}

%
\subsection{the space of items}

We suppose the minimum structure of the space of items
in recommender systems.
All recommender systems have the space of items to be recommended
and most of them have descriptions of each item's properties
({\it characteristics} or {\it item profiles})
to compare them to each other.
The most natural way to introduce such structure
is to use a vector space, i.e., the {\it vector space model}
(\cite{Jannach}).
We assume that our system has the vector space model.

Let $V = {\bf R}^N$ be the space of characteristics of items;
we express each item profile as a point
$x = ( x_i )_{i = 1, \dots, N} \in V$.
More precisely,
we have $N$ {\it features} to describe the properties,
and the $i$th feature of the characteristics $x$
has the weight $x_i \in {\bf R}$.
We naturally identify an item itself with the characteristics.

We also suppose that the system has a function
$\rho: V \times V \to [0, \infty)$
    to choose a set $N(x, \epsilon)$ of {\it near} items to a given item $x$
and a parameter $\epsilon$ by
\begin{equation}
    N(x, \epsilon) = N(x, \epsilon; V)
    = \{ y \in V: \rho(x, y) < \epsilon \}.
\end{equation}

For example, in a document recommender system,
the features may be words ({\it tokens} or {\it terms}),
the weights may be defined by TF-IDF algorithm,
and $\rho(\cdot, \cdot)$ may be the Euclidean metric
$ \rho (x, y) = \| x - y \|
 = \left( \sum_{i = 1}^N |x_i - y_i|^2 \right)^{1/2}$,
the cosine similarity
$\rho (x, y) = \sum_{i = 1}^N x_i y_i / (\| x \| \| y \|)$,
or others (\cite{Salton}, \cite{PazzaniBillsus}).

\subsection{the two-step model for recalling}

Consider a recommender system with the users and the items,
which repeats the following procedure.
The system recommends an item to each user,
who may take the recommendation (buy, read, etc.), or not.
It collects the information
which items each user took in the past
and analyses it to give a new recommendation to each user.

Now suppose that a user finds an interesting item,
which reminds the user some items recommended in the past.
We call the former item a {\it trigger}
and the latter {\it recalled} items.
We propose a function to pick up such {\it recalled} items
on behalf of the user.

We formulate this function as follows.
Let $R(u) ( \subset V)$
be the set of old recommendations to a user $u$
and $t \in V$ be a {\it trigger} recommended to the user $u$.
Our task is to pick up {\it recalled} items from the set $R(u)$
at the user's request.
We model this {\it recalling} as the following two steps.

First, we extend the {\it trigger} $t$ by the old recommendations $R(u)$
to get the {\it recall vector} $r = r(t, u) \in V$,
which is corresponding to an image in the user's mind
stimulated by $t$.
We explain precisely this part in the next subsection.
Second, we choose a set $\tilde{N}(t, \epsilon)$ of near items to
the {\it recall vector} $r$
by the method $\rho$ with (1), i.e.,
\begin{equation}
    \tilde{N}(t, \epsilon) = N(r(t, u), \epsilon; R(u)) =
        \{ y \in R(u): \rho(r, y) < \epsilon \}.
\end{equation}

\subsection{the feature relation matrices and the recall vectors}

In this subsection, we define the operation to generate {\it recall vectors}.
Roughly saying, we encode the co-occurrence information of the features
of the items in $R(u)$ into a matrix,
and we operate the matrix on the trigger to get the recall vector.

First, we prepare the {\it co-occurrence functions}
$c_{ij}(x): V \to {\bf R}$
that depend only on the two elements $x_i$ and $x_j$ of $x$.
Roughly saying, $c_{ij}(x)$ means how strongly
$j$th feature co-occurs when $i$th feature occurs in the item $x$.
The typical examples are a binary type
\[
 c_{ij} (x) = \left\{
     \begin{array}{cl}
        1 & \mbox{(if $x_i\neq 0$ and $x_j \neq 0$)},\\
        0 & \mbox{(otherwise),}
     \end{array}
 \right.
\]
or a proportional one
\[
 c_{ij} (x) = \left\{
     \begin{array}{cl}
        x_j / x_i  & \mbox{(if $x_i\neq 0$)},\\
        0 & \mbox{(otherwise)}.
     \end{array}
 \right.
\]

Second, we define the {\it feature relation matrix}
$(F_{ij})_{i,j = 1, \dots, N}$
as the conditional average of the co-occurrence function $c_{ij}$
when the $i$th feature occurs, i.e.,
\begin{equation}
    F_{ij} = F_{ij}(u) = \frac{1}{|C_i(u)|}\sum_{x \in C_i(u)} c_{ij}(x),
\end{equation}
where $C_i(u) = \{ x \in R(u) : x_i \neq 0 \}$
and $|C_i(u)|$ is the cardinality.

Now, we generate the {\it recall vector} of the trigger $t$ by
\begin{equation}
r = (r_j)_{j = 1 \dots, N}
= n (Ft)
= n \left( \left( \sum_{i=1}^N F_{ij} t_i
  \right)_{j = 1 \dots, N} \right),
\end{equation}
where $n(\cdot):V \to V$ is a normalising function
to adjust the recall vectors to the item space.
For example, $n(x) = x/\| x \|$,
or $n(x) = (s(x_i))_{i=1, \dots, N}$ with a sigmoid type function $s(\cdot)$,
etc.  according to the situation.

Note that this is a similar idea to the classical
neural networks for associative memory
(the self correlation model) originated independently
with Anderson(\cite{Anderson}), Kohonen, and Nakano
because the essence of our model
is to encode the co-occurrence information into a matrix
(though we do not necessarily subtract the diagonal elements).
We also remark that such co-occurrence matrices (or graphs)
are used to select critical features,
for example, in some algorithms for search engines (\cite{Ohsawa}).

\section{Discussions}

\subsection{regression analysis and computational cost}

Since $F_{ij}$ means relations between the $i$th and $j$th feature,
it is natural to consider it
as an estimation of a function $x_j = F_{ij}(x_i)$
and to generalise the linear operation $F_{ij} t_i$
to an functional $F_{ij}(t_i)$.
However, it is heavy to estimate $F_{ij}(\cdot)$ by regression analysis
or curve fitting with the data set $\{(x_i, x_j)\}_{x\in R(u)}$,
since we have $N^2$ functions to estimate.
Therefore our algorithm is a quick and simple substitution
for such heavy estimations.

Though the order of our algorithm also is
$O(|R(u)| N^2)$ times a cost of calculating $F_{ij}$,
we have the following advantages.
First, the cost for $F_{ij}$ is much smaller.
Second, in most of the real applications,
almost items have few features compared with $N$;
we can use the properties of sparse matrices
since our algorithm skips the weight $0$.
Third, we can update successively the matrices
to reduce the order
when each user gets a new recommendation.

\subsection{verification of the model}

Our model of {\it recalling} consists of the two steps:
to generate the {\it recall vector} as a mental image of a {\it trigger}
and to pick up the near items to it.
Though it seems difficult to verify the model itself,
we can test statistically the effectiveness
of choosing {\it recalled} items in a real recommender system.
For example, we can study the difference between
the {\it trigger} and the {\it recall vector}
to choose the {\it recalled} items, i.e.,
the difference between
$N(t, \epsilon)$ and $N(r(t, R(u)), \epsilon')$.
The following asymptotic co-occurrence function $c_{ij}(x)$
with a small parameter $\delta \geq 0$ should be useful for the comparison.
\[
 c_{ij} (x) = \left\{
     \begin{array}{cl}
        1 & \mbox{(if $i = j$ and $x_i\neq 0$)},\\
        \delta & \mbox{(if $i \neq j$, $x_i\neq 0$, and $x_j \neq 0$)},\\
        0 & \mbox{(otherwise)}.
     \end{array}
 \right.
\]

The {\it recall vector} $r$ should be
also fit for $\rho(r, x)$ to
measure how near each item $x$ is to $r$.
Though we ensure the adaptation by the normalising function $n(\cdot)$,
there is not necessarily a natural one;
we may need trial and error
with statistical tests in the real system.

\section{Conclusion}

\begin{itemize}
\item We proposed a new function of {\it recalling} in recommender systems,
    which picks up items {\it recalled} by a user
    from old recommendations to the user.
\item We proposed a two-step model for the {\it recalling}
    in recommender systems,
    which consists of generating the {\it recall vector}
    and choosing the neighbourhood in the space of old recommendations
    to the user.
\item We showed the algorithm to implement
    the function according to our model;
    the essence is to bring the idea of the classical neural networks
    for associative memory into item spaces of recommender systems.
\item Though statistical tests in a real system are left to the future,
    a practical implementation of our model should be possible.
\end{itemize}

\bibliographystyle{plain}

\end{document}